\documentclass[a4paper,11pt]{article}
\usepackage{jheppub} 
\usepackage[T1]{fontenc} 
\newcommand{\Trh}{T_\text{rh}}
\newcommand{\Tmax}{T_\text{max}}
\newcommand{\gs}{g_\star}
\newcommand{\gss}{g_{\star s}}
\newcommand{\sv}{\langle\sigma v\rangle}
\title{\boldmath Polynomial Inflation and Dark Matter}
\author[a]{Nicolás Bernal}
\author[b]{and Yong Xu}
\affiliation[a]{Centro de Investigaciones, Universidad Antonio Nariño\\
	Carrera 3 Este \# 47A-15, Bogotá, Colombia}
\affiliation[b]{\it Bethe Center for Theoretical Physics and Physikalisches
	Institut, Universit\"at Bonn\\
	Nussallee~12, 53115 Bonn, Germany}
\emailAdd{nicolas.bernal@uan.edu.co}
\emailAdd{yongxu@th.physik.uni-bonn.de}
\abstract{We present  a minimal UV complete framework to embed inflation and dark matter by extending the standard model with a singlet real scalar field (the inflaton) and a singlet fermonic field acting as dark matter. The inflaton features the most general renormalizable polynomial up to quartic order, which is flat due to the existence of a perturbed inflection-point, comfortably fitting CMB measurements. We also analyze (p)reheating  by considering the Higgs production via inflaton decay. In the early universe, dark matter can be generated by the mediation of gravitons or inflatons. However, the production via the direct decay of the inflatons dominates, making viable a large range of dark matter masses, from $\mathcal{O}(10^{-5})$~GeV to $\mathcal{O}(10^{11})$~GeV.}
\begin{document} 
	\begin{flushright}
		PI/UAN-2021-690FT
	\end{flushright}
	\maketitle
	\flushbottom
	%%%%%%%%%%%%%%%%%%%%%%%%%%%%%%%%%%%%%%%%%%%%%%%%%%%%%%%%%%%%%%%%%%%%%%%%%%%%%%%%%%%%%%%%%%%%%%%%%%%%%%%%%%%%%%%%%%%%%
	\section{Introduction}
	%%%%%%%%%%%%%%%%%%%%%%%%%%%%%%%%%%%%%%%%%%%%%%%%%%%%%%%%%%%%%%%%%%%%%%%%%%%%%%%%%%%%%%%%%%%%%%%%%%%%%%%%%%%%%%%%%%%%%%
	%%%%%%%%%%%%%%%%%%%%%%%%%%%%%%%%%%%%%%%
	% Current status for inflation
	%%%%%%%%%%%%%%%%%%%%%%%%%%%%%%%%%%%%%%%%
	Cosmic inflation is  an elegant paradigm to solve the major problems in standard cosmology~\cite{Starobinsky:1980te, Guth:1980zm, Linde:1981mu, Albrecht:1982wi}. The leading inflationary paradigm is the so-called slow roll (SR) inflation, in which the inflaton slowly rolls down its potential (see, e.g., Ref.~\cite{Lyth:2009zz}). During SR, a near scale-invariant power spectrum is generated due to vacuum fluctuations of the inflaton in a quasi-de Sitter background, in agreement with Cosmic Microwave Background (CMB) measurements~\cite{Akrami:2018odb}. Another notable prediction of inflation is the primordial gravitational wave spectrum due to (first-order) tensor perturbations of the metric, which is usually quantified by the so-called tensor-to-scalar ratio $r$. Latest experimental results favor $r < 0.061$~\cite{Akrami:2018odb, Ade:2018gkx} at the pivot scale $k_\star = 0.05$~Mpc$^{-1}$, which puts a strong constraint on the steepness of the inflationay potential. Indeed, the simplest monomial scenarios with $V(\phi) \propto \phi^p$  with $p \geq 1$ are disfavored~\cite{Akrami:2018odb}. This has motivated the study of flatter potentials see, e.g., Ref.~\cite{Martin:2013tda} for a comprehensive review.
	
	%%%%%%%%%%%%%%%%%%%%%%%%%%%%%%%%%%%%%%%%%%%%%%%%%%
	% WIMP 
	%%%%%%%%%%%%%%%%%%%%%%%%%%%%%%%%%%%%%%%%%%%%%%%%%%%%
	Moreover, observations indicate that the known baryonic matter could only account for $\sim 15\%$ of the total matter density~\cite{Aghanim:2018eyx}. The rest is unknown and called dark matter (DM)~\cite{Bertone:2016nfn}. Regarding its nature, weakly interacting massive particles (WIMPs) are one of the well-motivated candidates~\cite{Steigman:1984ac}. In the WIMP paradigm, DM with mass at the electroweak scale thermally freezes out from the SM plasma, reproducing the observed relic density $\Omega_{\rm DM} h^2 \simeq 0.12$~\cite{Aghanim:2018eyx}. WIMP scenarios call for new physics at the electroweak scale, which is remarkable since there are indeed a number of well-motivated SM extensions at such scale.
	However the null experimental results and strong observational constraints on the typical WIMP parameter space motivate quests beyond such paradigm~\cite{Arcadi:2017kky}. 
	
	%%%%%%%%%%%%%%%%%%%%%%%%%%%%%%%%%%%%%%%%%%%%%%%%%%%%%
	%FIMP
	%%%%%%%%%%%%%%%%%%%%%%%%%%%%%%%%%%%%%%%%%%%%%%%%%%%%%
	One alternative to the WIMP scenario is the so-called  feebly interacting massive particle (FIMP)~\cite{McDonald:2001vt, Choi:2005vq, Kusenko:2006rh, Petraki:2007gq, Hall:2009bx, Bernal:2017kxu}. In such case DM particles interact with the SM plasma so feebly that they never achieve chemical equilibrium with the thermal bath.
	%And DM particles  in such framework are generated by the freeze-in mechanism~\cite{Hall:2009bx} via interaction with the thermal bath. When  temperature of the thermal bath drops below mass of  DM particle, the abundance of DM are  frozen-in.
	A notable difference between FIMP and WIMP is that DM abundance via the former mechanism depends on initial conditions, namely, the initial density (typically assumed to be small) of DM particles, which can be produced during the cosmic heating era. 
	% Direct decay and gravitational production
	Besides, another possibility for the DM genesis is the direct decay of a heavy particle, for example the inflaton, a moduli field or the curvaton (see, e.g. Ref.~\cite{Baer:2014eja}). Additionally, DM particles could have also been produced via scattering of SM particles or inflatons mediated by the irreducible gravitational interaction~\cite{Garny:2015sjg, Tang:2016vch, Tang:2017hvq, Garny:2017kha, Bernal:2018qlk}. 
	
	%%%%%%%%%%%%%%%%%%%%%%%%%%%%%%%%%%%%%%%%%%%%%%%%%%
	% Unified Models in the market; what would be new?
	%%%%%%%%%%%%%%%%%%%%%%%%%%%%%%%%%%%%%%%%%%%%%%%%%
	In this work, we aim at solving inflation and DM in a unified way.
%	To that end one has to extend the minimal SM.
	The literature counts several proposals with extra singlet scalars acting simultaneously as the inflaton and DM (see, e.g., Refs.~\cite{Lerner:2009xg, Kahlhoefer:2015jma}), where nonminimal couplings between the inflaton and the Ricci scalar are usually introduced to flatten the inflaton potential. See also Refs.~\cite{Clark:2009dc, Khoze:2013uia, Almeida:2018oid, Bernal:2018hjm, Aravind:2015xst, Ballesteros:2016xej, Borah:2018rca,  Hamada:2014xka, Choubey:2017hsq, Cline:2020mdt, Tenkanen:2016twd, Abe:2020ldj} for similar works where nonminimal couplings are involved. However, inflaton could also couple minimally to gravity, for unified scenarios, see, e.g. the $\nu$MSM~\cite{Shaposhnikov:2006xi}, the NMSM~\cite{Davoudiasl:2004be}, SMART $U(1)_X$ \cite{Okada:2020cvq}, the  WIMPflation~\cite{Hooper:2018buz}, model with a single axion-like particle~\cite{Daido:2017wwb}, and extension with a  complex flavon  field \cite{Ema:2016ops}. For supersymmetric models see, e.g., Ref.~\cite{Allahverdi:2007wt}, inspired by the gauge invariant MSSM inflation~\cite{Allahverdi:2006iq}.
	
	%%%%%%%%%%%%%%%%%%%%%%%%%%%%%%%%%%%%%%%%%%%%%%%%%%%%
	% Why Polynomial inflation?
	%%%%%%%%%%%%%%%%%%%%%%%%%%%%%%%%%%%%%%%%%%%%%%%%%%%
	In this paper, we present a renormalizable and UV complete minimal framework to embed inflation and DM without the need of introducing nonminimal couplings to gravity. To that end, we consider a real scalar field $\phi$ acting as the inflaton. To avoid trans-Planckian issues~\cite{Bedroya:2019tba} and be consistent with the recent string swampland distance conjecture~\cite{Agrawal:2018own}, we will consider a small field inflation scenario, where the inflaton field value is sub-Planckian when the observable curvature perturbations were generated. In such case, the most general and renormalizable inflaton potential would be a polynomial up to a quartic term.
	
	The polynomial nature of the inflaton potential is a key difference from the usually considered monomial scenarios, which has been gaining interests since the 1990s~\cite{Hodges:1989dw, Destri:2007pv, Nakayama:2013jka, Aslanyan:2015hmi, Musoke:2017frr}. 
	%%%%%%%%%%%%%%%%%%%%%%%%%%%%%%%%%%%%%%%%%%%%%%%%
	% Why polynomial inflation interesting ?
	%%%%%%%%%%%%%%%%%%%%%%%%%%%%%%%%%%%%%%%%%%%%%%%%
	In the minimally coupled scenario, the flatness of a polynomial potential of quartic order is guaranteed by the existence of a (near) inflection-point~\cite{Drees:2021wgd}. It has been recently shown that such a polynomial inflection-point inflationary scenario fits easily consistent with current CMB observations with rather large parameter space~\cite{Drees:2021wgd}. Additionally, the inflationary predictions can be analyzed analytically even though the model includes several free parameters ~\cite{Drees:2021wgd}.
	A notable prediction is that the running of the spectral index $\alpha \sim \mathcal{O}(10^{-3})$, making this small field inflationary model testable in the near future~\cite{Munoz:2016owz}. Besides, the inflationary scale $H_I$ can be as low as $\sim 1\, \text{MeV}$ so that the cosmological moduli problem can be greatly alleviated~\cite{Coughlan:1983ci}, and additionally, the axion isocurvature bound can also be easily satisfied~\cite{Akrami:2018odb}. Since the reheating temperature can  exceed $10^{10}$~GeV, the standard thermal leptogenesis \cite{Davidson:2002qv} is then also possible . It is naturally therefore to extend the study presented in Ref.~\cite{Drees:2021wgd} in order to tackle the DM problem. 
	%%%%%%%%%%%%%%%%%%%%%%%%%%%%%%%%%%%%%%%%%%%%%%%%%%
	% Why fermion DM? --- it is simple:)  so Why not:D
	%%%%%%%%%%%%%%%%%%%%%%%%%%%%%%%%%%%%%%%%%%%%%%%%%%
	
	In this order of ideas, we consider an additional gauge singlet fermion field $\chi$ as the DM candidate. The possibility of singlet fermion DM was originally proposed in Ref.~\cite{Kim:2006af}, where DM interacts with the SM Higgs via dimension-5 operators. Later, the case where DM communicates with the SM via the mixing of a new singlet scalar field with the SM Higgs was studied~\cite{Kim:2008pp, Baek:2011aa, LopezHonorez:2012kv}.
	Here we show for the first time that the singlet scalar field can be identified as the inflaton. Moreover, a notable difference with previous works is that the mixing between the singlet scalar and SM Higgs has to be negligible small in order not to spoil the inflaton potential when Higgs loop corrections are included. 
	%%%%%%%%%%%%%%%%%%%%%%%%%%%%%%%%%%%%%%%%%%%%%%%%%%%%%%
	% How DM are produced? 
	%%%%%%%%%%%%%%%%%%%%%%%%%%%%%%%%%%%%%%%%%%%%%%%%%%%%%%
	As the inflaton directly couples to DM, the observed DM abundance can be generated via inflaton decay, or by the scattering of Higgs bosons mediated by the $s$-channel exchange of inflatons.
	Additionally, the irreducible gravitational interaction can also mediate the DM production, either by 2-to-2 annihilation of SM Higgs bosons, or of inflatons during reheating.
	
	%%%%%%%%%%%%%%%%%%%%%%%%%%%%%%%%%%%%%%%%%%%%%%%
	%Outline of paper.
	%%%%%%%%%%%%%%%%%%%%%%%%%%%%%%%%%%%%%%%%%%%%%%%%%
	The paper is organized as follows. In  Sec.~\ref{setup} we present the general setup for the model, whereas its inflationary predictions are reviewed in Sec.~\ref{inflation}. In Sec.~\ref{reheating} the heating dynamics is investigated. Then, in Sec.~\ref{dm} we study the production of the DM relic abundance in the early universe by considering direct decays of the inflaton, the UV freeze-in mechanism and gravitational scattering. Finally, in Sec.~\ref{summary}, we sum up our findings and end with some prospects for further extensions of the model.
	%%%%%%%%%%%%%%%%%%%%%%%%%%%%%%%%%%%%%%%%%%%%%%%%%%%%%%%%%%%%%%%%%%%%%%%%%%%%%%%%%%%%%%%%%%%%%
	\section{The Model Setup} \label{setup}
	Our model is defined to be the following action:
	\begin{equation}
	S = \int d^4x\, \sqrt{-g} \left(\mathcal{L}_{\rm EH} +  \mathcal{L}_{\rm SM} + \mathcal{L}_{\phi}+ \mathcal{L}_{\chi} + \mathcal{L}_{H\phi}\right),
	\end{equation}
	where $g$ is determinant of the Friedmann-Lemaître-Robertson-Walker metric, defined by $g_{\mu \nu} = \text{diag}(1,-a^2,-a^2,-a^2)$, with $a$ denoting the scale factor.
	Firstly, $\mathcal{L}_{\rm EH}$ denotes the Einstein-Hilbert Lagrangian given by
	\begin{equation}
	\mathcal{L}_{\rm EH}  = \frac{M_P^2}{2}\, R\,,
	\end{equation}
	where $R$ represents the Ricci scalar and $M_P \simeq 2.4 \times 10^{18}$~GeV is the reduced Planck mass.
	$\mathcal{L}_{\rm SM}$ describes the usual SM Lagrangian, and $\mathcal{L}_{\phi}$ corresponds to the inflaton sector
	\begin{equation}
	\mathcal{L}_{\phi} = \frac{1}{2} \partial_{\mu} \phi \partial^{\mu} \phi - V(\phi)\,,
	\end{equation}
	where for the real scalar inflaton potential $V(\phi)$ we are considering the most general renormalizable one\footnote{The linear term can be absorbed by shifting the field $\phi$. A similar potential was used in Ref.~\cite{DiChiara:2015euo} for the purpose of realizing hybrid inflation in the context of the cosmological relaxion model.}
	\begin{equation} \label{inflaton_potential}
		V(\phi)
		= b\, \phi^2 + c\, \phi^3 + d\, \phi^4.
	\end{equation}
	DM in our scenario is a gauge singlet Dirac fermion $\chi$ with a Lagrangian $\mathcal{L}_{\chi}$ given by
	\begin{equation}
	\mathcal{L_{\chi}} = i\, \bar{\chi}\, \gamma^{\mu}\, \partial_{\mu}\chi - m_{\chi}\, \bar{\chi}\, \chi - y_{\chi}\, \phi \bar{\chi}\, \chi \,,
	\end{equation}
	where $m_\chi$ corresponds to the DM mass.
	Finally, the term $\mathcal{L}_{H\phi}$  denotes interaction between the inflaton and the SM via the Higgs doublet $H$
	\begin{equation} \label{Vphih}
	\mathcal{L}_{H\phi} = -V(H,\phi) =  -\lambda_{12}\, \phi\, H^{\dagger} H - \frac{1}{2} \lambda_{22}\, \phi^2\, H^{\dagger} H  \,,
	\end{equation}
	which plays an important role to produce  SM particles and reheat the universe after inflation.
	%%%%%%%%%%%%%%%%%%%%%%%%%%%%%%%%%%%%%%%%%%%%%%%%%%%%%%%%%%%%%%%%%%%%%%%%%%%%%%%%%%%%%%%%%%%%%
	
	\section{Inflation} \label{inflation}
\begin{figure}[ht!]
		\centering
		\includegraphics[width=.6\paperwidth, keepaspectratio]{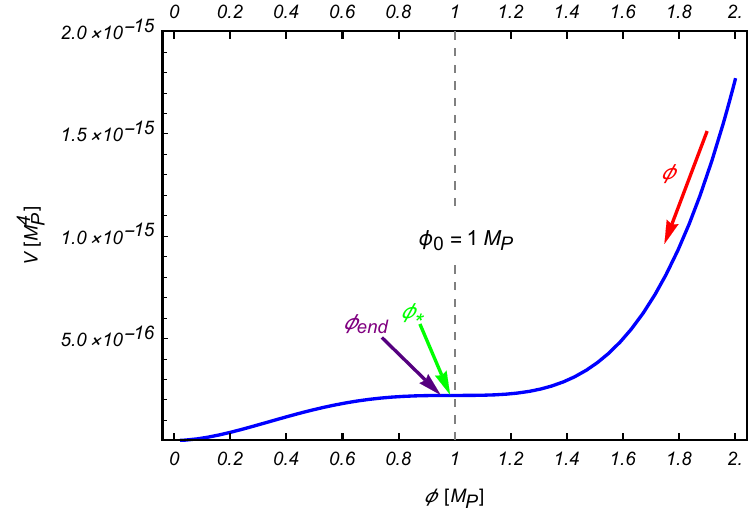}
		\caption {Inflaton potential with a near inflection point at $\phi = \phi_0 = M_P$ with $\beta>0$. The inflaton rolls from some high scale and then crosses $\phi_0$. Later when $\phi$ rolls to the regime with $\phi \sim \phi_\star$, the seeds (due to quantum fluctuation of $\phi$) for CMB anisotropies are generated. And finally inflation ends at $\phi_{\rm end}$.} 
		\label{potential}
	\end{figure}

	%%%%%%%%%%%%%%%%%%%%%%%%%%%%%%%%%%%%%%%%%%%%%%%%%%%%%%%%%%%%%%%%%%%%%%%%%%%%%%%%%%%%%%%%%%%%%
	In this section we briefly review SR inflation with focus on the polynomial inflection-point inflation formalism~\cite{Drees:2021wgd}.
	For $b = \frac{9}{32} \frac{c^2}{d}$, the potential in Eq.~\eqref{inflaton_potential} features an inflection-point at $\phi = \phi_0 \equiv -\frac{3}{8} \frac{c}{d}$, so one can reparametrize the potential as
	\begin{equation}  \label{inflaton_potential2}
		V(\phi) = d \left[2 \phi_0^2\, \phi^2 - \frac83 (1 - \beta)\, \phi_0\, \phi^3 + \phi^4\right],
	\end{equation}
	where a correction parameter $\beta$ is introduced to control the flatness configuration in the vicinity of $\phi_0$. In the case with $\beta = 0$, the exact inflection-point at $\phi_0$ is recovered, and if the inflaton excursion starts at $\phi > \phi_0$, it will get stuck at $\phi = \phi_0$.
	Alternatively, for the case with $\beta < 0$ a dangerous false vacuum for $\phi > \phi_0$ is generated.
	Here however, we will focus on the case $0<\beta \ll 1$, and on a small field inflationary scenario with $\phi_0 \lesssim M_P$. In Fig.~\ref{potential} the inflaton potential featuring an inflection point at $\phi_0 =M_P$ is shown.
	
	The SR parameters can be conveniently written by introducing the dimensionless parameter $\delta$ defined as
	\begin{equation} \label{delta}
	\delta \equiv 1 - \frac{\phi}{\phi_0}\,,
	\end{equation}
    and become~\cite{Drees:2021wgd}
	\begin{align}
		&\epsilon_V \equiv \frac{M_P^2}{2} \left(\frac{V^{\prime}}{V}\right)^2  \simeq  72 \left(2\beta  + \delta^2 \right)^2 \left(\frac{M_P}{\phi_0}\right)^2, \label{epsilon} \\
		&    \eta_V \equiv M_P^2\, \frac{V^{\prime \prime}}{V}  \simeq 24 \left(2 \beta - \delta\right) \left(\frac{M_P}{\phi_0}\right)^2, \label{eta}\\
		&\xi_V^2 \equiv M_P^4\, \frac{V^{\prime}\, V^{\prime \prime \prime}}{V^2} \simeq 288 \left(2\beta  + \delta^2\right) \left(\frac{M_P}{\phi_0}\right)^4 \label{xi}.
	\end{align}
	SR inflation requires $\epsilon_V, |\eta_V| \ll 1$. Since $\epsilon_V < |\eta_V|$ as can be seen from Eq.~(\ref{epsilon}) and Eq.~(\ref{eta}), SR inflation then ends at $\delta = \delta_\text{end}$, when $|\eta_V|=1$, which corresponds to
	\begin{equation}
	    \delta_\text{end} \simeq \frac{1}{24} \left(\frac{\phi_0}{M_P}\right)^2,
	\end{equation}
	where the fact that $\delta \gg \beta$ was used \cite{Drees:2021wgd}.
	The number of $e$-folds from $\phi_\star$ (defined at the pivot scale $k_\star = 0.05$~Mpc$^{-1}$) till the end of inflation is 
	\begin{equation}
	N_\star = \int^{\phi_\star}_{\phi_{\rm end}} \frac{1}{\sqrt{2\, \epsilon_V}} d\phi
	\simeq \frac{1}{12\sqrt{2\, \beta}} \left(\frac{\phi_0}{M_P}\right)^2 \left(\frac{\pi}{2} - \text{ArcTan}\left[\frac{\delta_\star}{\sqrt{2 \beta}}\right] \right ),
	\end{equation}
	where $\delta_\star = 1 - \frac{\phi_\star}{\phi_0}$.
	Additionally, the amplitude of scalar curvature perturbation, the spectral index and its running, and the tensor-to-scalar ratio are respectively given by
    \begin{align}
	\mathcal{P}_{\zeta} &= \frac{V}{24\pi^2\epsilon_V M_P^4} \simeq \frac{d}{5184 \pi^2\, (\delta_\star^2 + 2\beta)^2} \left(\frac{\phi_0}{M_P}\right)^6,\\
	n_s &= 1- 6\epsilon_V + 2\eta_V \simeq 1- 48\, \delta_\star \left(\frac{M_P}{\phi_0}\right)^2,\\
	\alpha &= 16\, \epsilon_V\, \eta_V - 24\, \epsilon_V^2 -2\, \xi_V^2 \simeq - 576\, \left(2 \beta +\delta_\star^2\right) \left(\frac{M_P}{\phi_0}\right)^4,\\
	r &= 16\, \epsilon_V \simeq 1152\, \left(2\beta + \delta_\star^2\right)^2\, \left(\frac{M_P}{\phi_0}\right)^2.
	\end{align}
	CMB measurements from the Planck collaboration~\cite{Akrami:2018odb} set strong constraints on the scalar power spectrum ($\mathcal{P}_{\zeta} = \left(2.1 \pm 0.1 \right) \times 10^{-9}$ and $n_s = 0.9649 \pm 0.0042$), which for $N_\star = 65$ $e$-folds imply~\cite{Drees:2021wgd}
	\begin{align}
	\delta_\star  &\simeq 7.31 \times 10^{-4} (\phi_0/M_P)^2\,, \label{pa1}\\
	\beta &\simeq 9.73 \times  10^{-7} (\phi_0/M_P)^4\,,\label{pa2} \\
	d &\simeq 6.61 \times  10^{-16} (\phi_0/M_P)^2\,, \label{pa3}
	\end{align}
	from which one can also see that $\delta \gg \beta$ as has assumed earlier. With these model parameters, the tensor-to-scalar ratio becomes
	\begin{equation} \label{r}
	r \simeq 7.09 \times  10^{-9} (\phi_0/M_P)^6,
	\end{equation}
	well bellow the present bound ($r < 0.061$~\cite{Akrami:2018odb}) and the expected sensitivity of future detectors.
	However, the prediction for the running of spectral index is $\alpha \simeq -1.43 \times 10^{-3}$, independently of $\phi_0$, in agreement with current bounds $\alpha = -0.0045 \pm 0.0067$~\cite{Akrami:2018odb}, and potentially testable with future CMB measurements combined with improved understanding of structures at small scale, in particular the Lyman-$\alpha$ forest~\cite{Munoz:2016owz}.
	Finally, we note that the inflaton mass $m_\phi$ is given by
    \begin{equation} \label{mass}
        m_\phi = 2\sqrt{d}\, \phi_0 \simeq 5.14 \times 10^{-8}\, \phi_0^2/M_P\,.
    \end{equation}
    As it will be seen in next section, $\phi_0$ spans in the range $\sim 10^{-5}~M_P$ to $M_P$, and therefore the inflaton mass $10^2$~GeV $\lesssim m_\phi \lesssim 10^{11}$~GeV.
	
	%%%%%%%%%%%%%%%%%%%%%%%%%%%%%%%%%%%%%%%%%%%%%%%%%%%%%%%%%%%%%%%%%%%%%%%%%%%%%%%%%%
	\section{Reheating} \label{reheating}
	After the end of inflation, the universe is heated via the interactions between the inflaton and the SM Higgs.
	Before analyzing the perturbative decay of inflaton, we will first briefly discuss the effect of possible nonperturbative preheating~\cite{Kofman:1997yn}. 
	%%%%%%%%%%%%%%%%%%%%%%%%%%%%%%%%%%%%%%%%%%%%%%%%%%%%%%%%%%%%%%%%%%%%%%%%%%%%%%%%%%
	\subsection{Remarks on Preheating}
	%%%%%%%%%%%%%%%%%%%%%%%%%%%%%%%%%%%%%%%%%%%%%%%%%%%%%%%%%%%%%%%%%%%%%%%%%%%%%%%%
	Preheating corresponds to the nonperturbative particle production due to parametric resonance, for a review see, e.g., Ref.~\cite{Lozanov:2019jxc}.
	In this scenario, preheating could proceed via the trilinear or the quartic coupling of the inflaton with the SM Higgs, described in Eq.~\eqref{Vphih}.
	
	Let us first focus on the preheating proceeding via the trilinear coupling $\lambda_{12}\, \phi\, H^\dagger H$.
	Such a term gives rise to a tachyonic effective mass square contribution ($\propto \lambda_{12}\, \phi$) for the Higgs field once the inflaton crosses zero and becomes negative during the oscillation, and therefore preheating tends to become efficient due to tachyonic resonance~\cite{Dufaux:2006ee}. However, the Higgs self-coupling $\lambda_H\, |H^\dagger H|^2$ induces a positive effective mass $\propto \lambda_H \langle  H^2\rangle$ (where $\langle H^2\rangle$ denotes the variance of the produced Higgs field), which counteracts on the tachyonic effective  mass. Such backreaction effect blocks the energy transfer from the inflaton, making preheating inefficient.
	Detailed lattice simulations show that trilinear interactions also speed up thermalization, having a time scale much shorter than the lifetime of the inflaton~\cite{Dufaux:2006ee}. 
		
	Alternatively, preheating could also occur via the quartic coupling $\lambda_{22}\, \phi^2\, H^\dagger H$, and is efficient if $\lambda_{22} \gtrsim \mathcal{O}(10^{-8})$~\cite{Kofman:1997yn}.
	However, to guarantee the radiative stability of the inflaton potential, one has to impose that $\lambda_{22} \lesssim \mathcal{O}(10^{-10})$~\cite{Drees:2021wgd}, making this channel negligible.
	Having shown that preheating effects are subdominant, we now focus on reheating due to the perturbative decay of the inflaton.

	%%%%%%%%%%%%%%%%%%%%%%%%%%%%%%%%%%%%%%%%%%%%%%%%%%%%%%%%%%%%%%%%%%%%%%%%%%%%%%%%%
	\subsection{Perturbative Decay of the Inflaton} 
	%%%%%%%%%%%%%%%%%%%%%%%%%%%%%%%%%%%%%%%%%%%%%%%%%%%%%%%%%%%%%%%%%%%%%%%%%%%%%%%%%%
	The inflaton can decay into SM Higgs bosons or DM particles, with decay rates given by
	\begin{align}
	\Gamma_{\phi \to H^{\dagger} H}  &\simeq \frac{\lambda_{12}^2}{8\pi\, m_\phi}\,,\\
	\Gamma_{\phi \to \bar\chi \chi}  &\simeq \frac{y_\chi^2\, m_\phi}{8\pi}\,,
	\end{align}
	and therefore the total decay width $\Gamma \equiv \Gamma_{\phi \to H^{\dagger} H} + \Gamma_{\phi \to \bar\chi \chi} \simeq \Gamma_{\phi \to H^{\dagger} H}$, as it has to dominantly decay into SM states. We define Br to be  the branching ratio into a couple of DM particles, i.e.,
	\begin{equation} \label{br}
	\text{Br} \equiv \frac{\Gamma_{\phi \to \bar{\chi}\chi} }{\Gamma_{\phi \to \bar{\chi}\chi} + 	\Gamma_{\phi \to H^{\dagger} H}} \simeq  \frac{\Gamma_{\phi \to \bar{\chi}\chi} }{\Gamma_{\phi \to H^{\dagger} H}} \simeq  2.6\times 10^{-15}\, \frac{\phi_0^4\, y_{\chi}^2}{M_P^2\, \lambda_{12}^2}\,. 
	\end{equation}

	The reheating temperature $\Trh$, denoting the onset of the radiation-dominated era, can be defined as $H(\Trh) = \frac23 \Gamma$, and therefore
	\begin{equation}\label{tre}
	\Trh = \sqrt{\frac{2}{\pi}} \left(\frac{10}{\gs}\right)^{1/4} \sqrt{M_P\, \Gamma}
	\simeq  3.9 \times 10^2 \left(\frac{\lambda_{12}}{\phi_0}\right) M_P\,,
	\end{equation}
	where $\gs(T)$ denotes the number of relativistic degrees of freedom contributing to the SM energy density $\rho_R$, with $\gs = 106.75$ for temperatures much higher than the $t$-quark mass.
	In order to guarantee a successful BBN $\Trh \gtrsim 4$~MeV~\cite{Sarkar:1995dd, Kawasaki:2000en, Hannestad:2004px, DeBernardis:2008zz, deSalas:2015glj}.
	Additionally, the upper limit on the inflationary scale $H_I \leq 2.5 \times 10^{-5}M_P$~\cite{Akrami:2018odb} allows to extract the upper bound $\Trh \lesssim 7 \times 10^{15}$~GeV.
	Therefore, one has that
	\begin{equation} \label{constrain1}
	2.4 \times 10^{-24} \lesssim \frac{\lambda_{12}}{\phi_0} \lesssim 7.5 \times 10^{-6}.
	\end{equation}
	The ratio $\lambda_{12}/\phi_0$ can be further constrained by ensuring that the inflaton potential is radiatively stable near inflection-point against Higgs loop corrections during inflation, which translates into~\cite{Drees:2021wgd}
    \begin{equation} \label{constrain2}
	\left|\left(\frac{\lambda_{12}}{\phi_0}\right)^2 \ln\left(\frac{\lambda_{12}}{\phi_0}\right) - \left(\frac{\lambda_{12}}{\phi_0}\right)^2\right| < 64\pi^2 d\, \beta \simeq 6.1 \times 10^{-19} \left(\frac{\phi_0}{M_P}\right)^6.
	\end{equation}
    Similarly, in order not to spoil the flatness of  potential near inflection-point by including DM loops, $y_{\chi}$ has to satisfy~\cite{Drees:2021wgd}
	\begin{equation} \label{ychibound}
	\left|y_{\chi}^4  - 3 y_{\chi}^4 \ln\left(y_{\chi}^2\right)\right| < 6.1 \times 10^{-19} \left(\frac{\phi_0}{M_P}\right)^6.
	\end{equation}

	While reheating is commonly approximated as an instantaneous event, the decay of the inflaton is a continuous process during which the bath temperature may rise to a value $\Tmax$ greater than $\Trh$~\cite{Giudice:2000ex}.
	Taking into account that the inflationary scale $H_I$ is given by
	\begin{equation} \label{hinf}
	    H_I \simeq \sqrt{\frac{V(\phi_0)}{3M_P^2}} \simeq 8.6 \times 10^{-9} \frac{\phi_0^3}{M_P^2}\,,
	\end{equation}
	the ratio $\Tmax/\Trh$ can be expressed as~\cite{Bernal:2019mhf}
	\begin{equation} \label{Tem_ratio}
	    \frac{\Tmax}{\Trh} = \left(\frac38\right)^{2/5} \left(\frac{H_I}{H(\Trh)}\right)^{1/4} \simeq 4.8 \times 10^{-3} \left(\frac{\phi_0^3}{M_P\, \Trh^2}\right)^{1/4}.
	\end{equation}

	%%%%%%%%%%%%%%%%%%%%%%%%%%%%%%%%%%%%%%%%%%%%%%%%%%%%%%
    \begin{figure}[t!]
        \def\sepf{0.50}
    	\centering
    	\includegraphics[scale=\sepf]{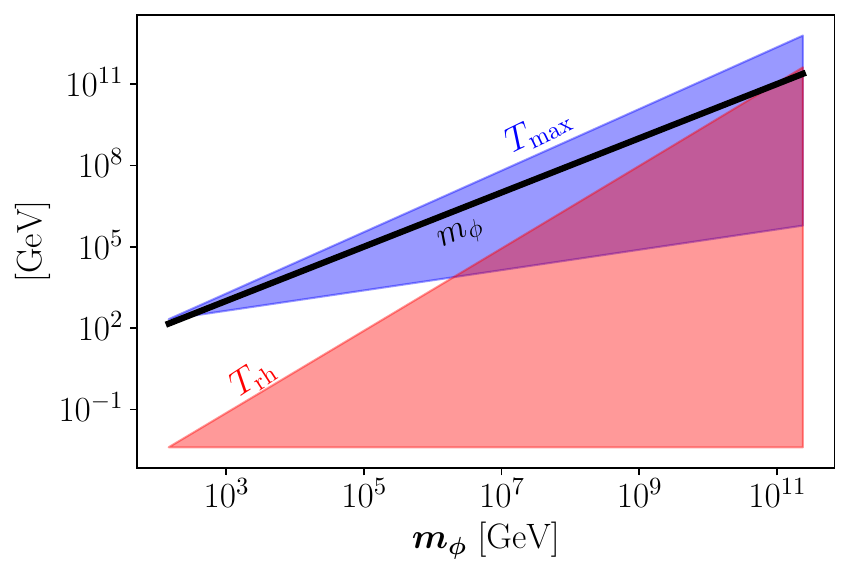}
    	\includegraphics[scale=\sepf]{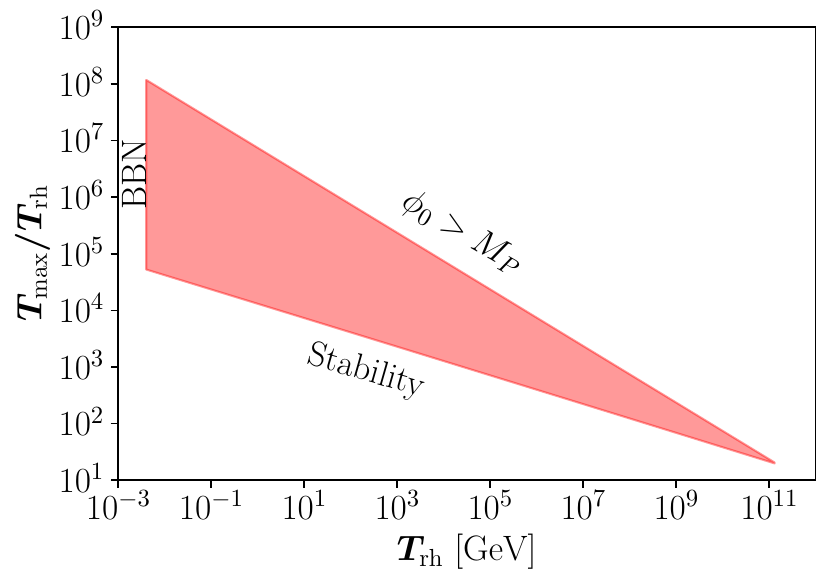}
        \caption{Left panel: Allowed ranges for $\Trh$ and $\Tmax$ as a function of the inflaton mass $m_\phi$.
        Right panel: Allowed range for the ratio $\Tmax/\Trh$ as a function of $\Trh$.}
    	\label{fig:RH}
    \end{figure} 
    %%%%%%%%%%%%%%%%%%%%%%%%%%%%%%%%%%%%%%%%%%%%%%%%%%%%%%
    It turns out that to satisfy the previously listed constraints  (BBN and stability) corresponding to Eqs.~\eqref{constrain1} and~\eqref{constrain2}, the lower bound for $\phi_0$ is $\phi_0 \gtrsim 3.0 \times 10^{-5}\, M_P$ \cite{Drees:2021wgd}, which together with $\phi_0 \lesssim M_P$ imply $10^2$~GeV $\lesssim m_\phi \lesssim 10^{11}$~GeV.
    Additionally, the reheating temperature spans over a large range, from $T_\text{BBN}$ to $\sim 10^{11}$~GeV as shown by red band in the left panel of Fig.~\ref{fig:RH}, whereas $\Tmax$ (blue band) ranges from $\sim 10^{2}$~GeV  to $\sim 10^{12}$~GeV. Inflaton mass is shown by the black line, which spanning from $\sim 10^{2}$~GeV  to $\sim 10^{11}$~GeV. The ratio $\Tmax/\Trh \subset [10^1,\, 10^8]$~GeV as shown in the right panel of Fig.~\ref{fig:RH}.
    %%%%%%%%%%%%%%%%%%%%%%%%%%%%%%%%%%%%%%%%%%%%%%%%%%%%%%
	%%%%%%%%%%%%%%%%%%%%%%%%%%%%%%%%%%%%%%%%%%%%%%%%%%%%%%%%%%%%%%%%%%%%%%%%%%%%%%%%%%%%%%%%%%%%%%%%%%%%%%%%%%%%%%%%%%%
	\section{Dark Matter Production and Relic Density} \label{dm}
\begin{figure}[ht!]
		\centering
		\includegraphics[scale=0.4]{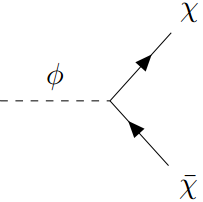}\\
		\includegraphics[scale=0.4]{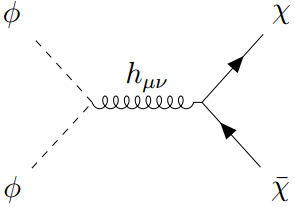}\qquad
		\includegraphics[scale=0.4]{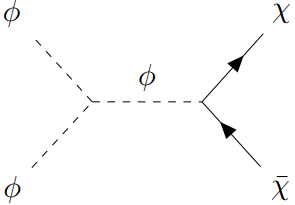}\qquad
		\includegraphics[scale=0.4]{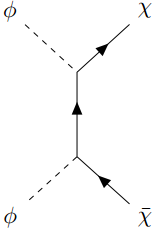}\\
		\includegraphics[scale=0.4]{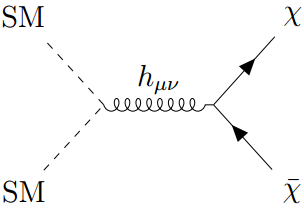}\qquad
		\includegraphics[scale=0.4]{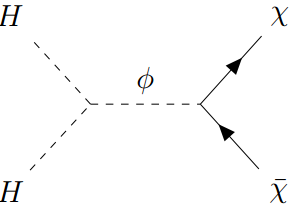}
		\caption {Feynman  diagrams for the different DM productions channels described in the text.
		The three rows correspond to $i)$ the decay of the inflaton, $ii)$ inflaton scatterings, and $iii)$ the scattering of SM particles.} 
		\label{diagam}
	\end{figure}

	%%%%%%%%%%%%%%%%%%%%%%%%%%%%%%%%%%%%%%%%%%%%%%%%%%%%%%%%%%%%%%%%%%%%%%%%%%%%%%%%%%%%%%%%%%%%%%%%%%%%%%%%%%%%%%%%%%%%
	In the early universe, DM can be produced via different processes.
	The main contributions come from $i)$ the direct decay of inflatons, $ii)$ 2-to-2 annihilations of inflatons during the reheating era (mediated by the $s$-channel exchange of inflatons or gravitons, and the $t$- and $u$-channel exchange of DM), and $iii)$ 2-to-2 annihilations of SM particles mediated by gravitons and inflatons.
	The corresponding Feynman diagrams are collected in Fig.~\ref{diagam}.
	%%%%%%%%%%%%%%%%%%%%%%%%%%%%%%%%%%%%%%%%%%%%%%%%%%%%%%%%%%
	%%%%%%%%%%%%%%%%%%%%%%%%%%%%%%%%%%%%%%%%%%%%%%%%%%%%%%%%%%
	
	The evolution of the DM number density $n$ is given by the Boltzmann equation
	\begin{equation}\label{eq:DMBE01}
	    \frac{dn}{dt} + 3H\,n  = \gamma\,, %-\langle\sigma v\rangle\left(n^2-n_\text{eq}^2\right)+2\,\text{Br}\,\Gamma\,\frac{\rho_\phi}{m_\phi}\,.
	\end{equation}
	where $\gamma$ corresponds to the DM production rate density and $H$ to the Hubble expansion rate given by $H^2 = (\rho_R + \rho_\phi)/(3 M_P^2)$.
	The evolution of the inflaton and SM radiation energy densities ($\rho_\phi$ and $\rho_R$, respectively) can be tracked via the set of Boltzmann equations
	\begin{align}
	    &\frac{d\rho_\phi}{dt} + 3 H\, \rho_\phi = - \Gamma\, \rho_\phi\,,\\
	    &\frac{d\rho_R}{dt} + 4 H\, \rho_R = + \Gamma\, \rho_\phi \left(1 - \text{Br}\right),
	\end{align}
	where factor $(1 - \text{Br}) \simeq 1$ is the fraction of inflaton energy density that goes into SM radiation.
	
	During reheating (i.e., for $\Trh < T < \Tmax$), the total energy density of the universe is dominated by inflatons.
	Taking into account that $T \propto a^{-3/8}$ and that the inflaton energy density scales as nonrelativistic matter,
	\begin{equation}
	    \rho_\phi(T) = \frac{\pi^2\, \gs}{30}\, \frac{T^8}{\Trh^4}\,,
	\end{equation}
	where $\rho_\phi(\Trh) = \rho_R(\Trh)$ was assumed.
	Therefore, the Hubble parameter in an inflaton-dominated universe takes the form
	\begin{equation}
	    H(T) = \frac{\pi}{3} \sqrt{\frac{\gs}{10}} \frac{T^4}{M_P\, \Trh^2}\,.
	\end{equation}
	The evolution of the DM number density given in Eq.~\eqref{eq:DMBE01} can be recasted in terms of the comoving number density $N \equiv n\, a^3$, where $a$ corresponds to the scale factor, considering the fact that during the reheating era the SM entropy density is not conserved due to the inflaton decay, and therefore
	\begin{equation} \label{eq:BEduring}
	    \frac{dN}{dT} = -\frac{8}{\pi} \sqrt{\frac{10}{\gs}} \frac{M_P\, \Trh^{10}}{T^{13}} a^3(\Trh)\, \gamma\,.
	\end{equation}
	
	However, after reheating (i.e., for $T < \Trh$), the universe is dominated by SM radiation and Eq.~\eqref{eq:DMBE01} can be recasted as a function of the DM yield $Y(T) \equiv n(T)/s(T)$, defined as a function of the SM entropy density $s(T) \equiv \frac{2\pi^2}{45} \gss T^3$, with $\gss(T)$ being the number of relativistic degrees of freedom contributing to the SM entropy~\cite{Drees:2015exa}.
	Therefore, Eq.~\eqref{eq:DMBE01} becomes
	\begin{equation} \label{eq:BEafter}
	    \frac{dY}{dT} = - \frac{135}{2\pi^3\, \gss} \sqrt{\frac{10}{\gs}}\, \frac{M_P}{T^6}\, \gamma\,.
	\end{equation}

	Finally, let us note that to reproduce the observed DM energy density $\Omega_\text{DM} h^2\simeq 0.12$~\cite{Aghanim:2018eyx}, the DM yield has been fixed such that 
	\begin{equation}
	m_\chi\,Y_0 = \Omega_\chi h^2 \, \frac{1}{s_0}\,\frac{\rho_c}{h^2} \simeq 4.3 \times 10^{-10}~\text{GeV}\,,
	\end{equation}
	with $\rho_c \simeq 1.05 \times 10^{-5} \, h^2$~GeV/cm$^3$ the critical energy density and $s_0\simeq 2.9\times 10^3$~cm$^{-3}$ the present entropy density~\cite{Zyla:2020zbs}.
	
	In the present scenario, where DM only communicates to the SM via the inflaton and gravity, the DM relic abundance could be produced via a number of channels.
	These different processes will be separately studied in the following subsections.
	
	%%%%%%%%%%%%%%%%%%%%%%%%%%%%%%%%%%%%%%%%%%%%%%%%%%%%%%%%%%%%%%%%%%%%%%%%%%%%%%%%%%%%%%%%%%%%%%%%%%%%%%%%%%%%%%%%%%%%
	\subsection{Inflaton Decay} \label{dir_decay}
	%%%%%%%%%%%%%%%%%%%%%%%%%%%%%%%%%%%%%%%%%%%%%%%%%%%%%%%%%%%%%%%%%%%%%%%%%%%%%%%%%%%%%%%%%%%%%%%%%%%%%%%%%%%%%%%%%%%%
    Due to the trilinear coupling $y_\chi\, \phi\, \bar{\chi} \chi$, DM  can be produced by direct decays of the inflaton.
    In this case, corresponding to the first row of Fig.~\ref{diagam}, the decay rate density is given by
	\begin{equation}
	    \gamma = 2\, \text{Br}\, \Gamma\, \frac{\rho_\phi}{m_\phi}\,.
	\end{equation}
	During reheating, i.e. while $\Tmax > T > \Trh$, Eq.~\eqref{eq:BEduring} can be solved analytically as
	\begin{equation}
	    N \simeq \frac{2\pi\, \gs}{15} \sqrt{\frac{10}{\gs}} \frac{M_P\, \Trh^2}{m_\phi} a^3(\Trh)\, \text{Br}\, \Gamma\,,
    \end{equation}
    which implies that
	\begin{equation} \label{y0total}
	    Y_0 = \frac{N(\Trh)}{s(\Trh)\, a^3(\Trh)} \simeq \frac{3}{\pi} \frac{\gs}{\gss} \sqrt{\frac{10}{\gs}} \frac{M_P\, \Gamma}{m_\phi\, \Trh} \text{Br} \simeq \frac32 \frac{\gs}{\gss} \frac{\Trh}{m_\phi} \text{Br}\,.
    \end{equation}
	Therefore to produce the whole observed DM abundance, one needs:
	\begin{equation} \label{ytm}
	    y_\chi \simeq 1.2 \times 10^{-13} \sqrt{\frac{\Trh}{m_\chi}}\,.
	\end{equation}
	%%%%%%%%%%%%%%%%%%%%%%%%%%%%%%%%%%%%%%%%%%%%%%%%%%%%%%
	\begin{figure}
        \def\sepf{0.50}
    	\centering
    	\includegraphics[scale=\sepf]{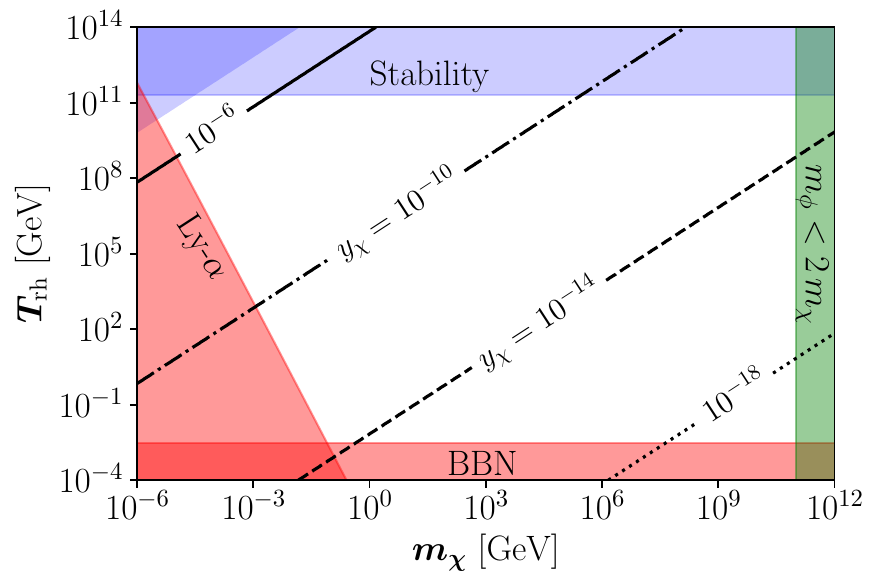}
		\caption{Yukawa coupling $y_\chi$ required to reproduce the whole observed DM abundance via the direct decay of the inflaton.
		The colored bands correspond to the different constraints described in the text.  The white region is the allowed parameter space with $10^{-21}\lesssim \text{Br} \lesssim 10^{-4}$ and  $\mathcal{O}(10^{-5})~\text{GeV} \lesssim m_\chi \lesssim \mathcal{O}(10^{11})~\text{GeV}$.}
	\label{fig:decay}
	\end{figure} 
	%%%%%%%%%%%%%%%%%%%%%%%%%%%%%%%%%%%%%%%%%%%%%%%%%%%%%%
    The Yukawa coupling $y_\chi$ required to reproduce the whole observed DM abundance via the direct decay of the inflaton is shown in Fig.~\ref{fig:decay}.
    The colored bands correspond to different constraints discussed: BBN (left side of Eq.~\eqref{constrain1}), radiative stability of inflaton potential (Eqs.~\eqref{constrain2} and~\eqref{ychibound}), Lyman-$\alpha$ (Eq.~\eqref{ly_mass}), and the kinematical threshold $m_\phi > 2\, m_\chi$ (cf. Eq.~\eqref{mass}).  The white area corresponds to the allowed parameter space, with branching ratio $10^{-21} \lesssim \text{Br} \lesssim 10^{-4}$; the upper bound on the branching fraction comes from the Lyman-$\alpha$ bound as shown in Eq.~\eqref{br_bound}, while the lower bound corresponds to the case with $\Trh \simeq 1.2 \times 10^{11}$~GeV and $m_\chi \simeq 6 \times 10^{10}$~GeV. The DM mass in the allowed parameter space can span from $\mathcal{O}(10^{-5})$~GeV to $\mathcal{O}(10^{11})$~GeV. Here we would like to stress that the allowed parameter space, namely the while region corresponds to the combinations of the bounds mentioned above with all possible $\phi_0$ or equivalently $m_\phi$.

	%%%%%%%%%%%%%%%%%%%%%%%%%%%%%%%%%%%%%%%%%%%%%%%%%%%%%%%%%%%%%%%%%%%%%%%%%%%%%%%%%%%%%%%%%%%%%%%%%%%%%%%%%%%%%%%%%%%%
	\subsection{Inflaton Scattering}
	%%%%%%%%%%%%%%%%%%%%%%%%%%%%%%%%%%%%%%%%%%%%%%%%%%%%%%%%%%%%%%%%%%%%%%%%%%%%%%%%%%%%%%%%%%%%%%%%%%%%%%%%%%%%%%%%%%%%
	Alternatively, DM can be produced during reheating by 2-to-2 scatterings of inflatons mediated by the $s$-channel exchange of gravitons or inflatons, and the $t$- and $u$-channel exchange of a DM particle, as shown in the second row of Fig.~\ref{diagam}. However the second and third processes are {\it always} sub-dominant compared to the direct decay due to a further coupling suppression.
	We will therefore focus on the gravitational channel, which might dominate in particular if the branching ratio Br is suppressed.
	%Note that for $\text{Br}< 10^{-21}$, {\it no} enough DM relics can be generated within the parameter space shown in  Fig.~\ref{fig:decay}.

    In the case where the graviton mediation dominates, the interaction rate density for DM production out of nonrelativistic inflatons reads~\cite{Mambrini:2021zpp, Bernal:2021kaj, Barman:2021ugy}
    \begin{equation}
        \gamma = \frac{\pi^3\, \gs^2}{3686400} \frac{T^{16}}{M_P^4\, \Trh^8} \frac{m_{\chi}^2}{m_\phi^2} \left(1-\frac{m_{\chi}^2}{m_\phi^2}\right)^{3/2}.
    \end{equation}
    The DM yield at the end of reheating can be analytically computed with Eq.~\eqref{eq:BEafter} as
	\begin{align}
	    Y_0 &\simeq \frac{\gs^2}{81920\gss} \sqrt{\frac{10}{\gs}} \left(\frac{\Trh}{M_P}\right)^3 \left[\left(\frac{\Tmax}{\Trh}\right)^4 - 1 \right] \frac{m_{\chi}^2}{m_\phi^2} \left(1-\frac{m_{\chi}^2}{m_\phi^2}\right)^{3/2}\nonumber\\
	    &\simeq 1.8 \times 10^{-2} \frac{\Trh\, m_\chi^2}{M_P^{5/2}\, m_\phi^{1/2}}\left(1-\frac{m_{\chi}^2}{m_\phi^2}\right)^{3/2}\,.
    \end{align}
    However taking into account the stability constraint  presented in Eq.~\eqref{constrain2} on the reheating temperature, i.e. $\Trh \lesssim 10^{11}$~GeV, it follows that DM production via inflaton scatterings  can contribute at most to few percent of the total DM abundance.
    
	%%%%%%%%%%%%%%%%%%%%%%%%%%%%%%%%%%%%%%%%%%%%%%%%%%%%%%%%%%%%%%%%%%%%%%%%%%%%%%%%%%%%%%%%%%%%%%%%%%%%%%%%%%%%%%%%%%%%
	\subsection{SM Scattering}
	Finally, DM could also be produced by the scattering of SM particles, as shown by the third row of Fig.~\ref{diagam}.
	This channel, corresponding to the UV freeze-in, can be mediated by the $s$-channel exchange of gravitons or inflatons, and is presented in the following.
	
	%%%%%%%%%%%%%%%%%%%%%%%%%%%%%%%%%%%
	\subsubsection{Graviton Mediation}
	Here we investigate  production of DM via the scattering of SM particles with gravitons acting as the mediator. This gravitational production mechanism is unavoidable due to the universal couplings between the metric and the energy-momentum tensor $\sim g_{\mu \nu}\, h^{\mu \nu}$.%
	\footnote{Two comments are in order: First, one could also conceive scenarios where gravity and another portal are effective, see, e.g., Refs.~\cite{Chianese:2020yjo, Chianese:2020khl, Gondolo:2020uqv, Bernal:2020bjf, Bernal:2021akf}.
    Second, the gravitational production can be enhanced in scenarios with extra dimensions, see e.g., Refs.~\cite{Lee:2013bua, Rueter:2017nbk,  Rizzo:2018joy, Brax:2019koq, Folgado:2019sgz,  Folgado:2019gie, Bernal:2020fvw, Bernal:2020yqg}.}
    The interaction rate density can be written as~\cite{Garny:2015sjg, Tang:2017hvq, Bernal:2018qlk}
	\begin{equation}
	\gamma(T) = \alpha\, \frac{T^8}{M_P^4}\,,
	\end{equation}
	where $\alpha \simeq 1.1 \times 10^{-3}$.
	The DM abundance produced after reheating is given by integrating Eq.~\eqref{eq:BEafter}, and therefore
	\begin{equation} \label{smg1}
	Y_0 = \frac{45 \alpha}{2\pi^3 \gss} \sqrt{\frac{10}{\gs}} \left(\frac{\Trh}{M_P}\right)^3, \qquad \text{ for } m_\chi \ll \Trh\,.
	\end{equation}

    As in the previous case, away from the instantaneous decay approximation, the DM yield is only boosted by a small factor of order $\mathcal{O}(1)$, as $\gamma(T) \propto T^8$.
	In the case where DM is heavier than the reheating temperature (but still lighter than $\Tmax$), one should compute the DM abundance during the reheating era.
	Equation~\eqref{eq:BEduring} yields therefore
	\begin{equation} \label{smg2}
	Y_0 = \frac{45 \alpha}{2\pi^3 \gss} \sqrt{\frac{10}{\gs}} \frac{\Trh^7}{M_P^3\, m_\chi^4}\,, \qquad \text{ for } m_\chi \gg \Trh\,.
	\end{equation}
	
	However, in the present scenario this DM production mediated by the exchange of gravitons cannot generate enough DM relics as the reheating temperature of the universe is $\Trh \lesssim 10^{11}$~GeV in order to not spoil the flatness of the inflaton potential by radiative corrections.

    We now close this section with a few remarks  regarding  another purely gravitational production channel. Indeed any  massive particle (which violates conformal invariance), and in particular the DM field, can be generated  due to the effect of time variations of the background metric, in particular during inflaton oscillations~\cite{Ford:1986sy, Kuzmin:1998kk, Chung:1998zb, Chung:2001cb, Chung:2004nh, Ema:2015dka, Ema:2016hlw, Ema:2018ucl, Ema:2019yrd}. Through this mechanism DM particles with mass $m_\chi \lesssim H_I$ can be produced with typical number density  $n_\chi \sim  H_I^3$~\cite{Chung:2001cb}, and it is particularly relevant for $m_\chi \simeq H_I$~\cite{Garny:2015sjg, Garny:2017kha}. Via this mechanism, the DM relic density scales like $\Omega_\chi h^2 \sim \left( m_\chi/10^{11}~\text{GeV}\right)^2 (\Trh/10^9~\text{GeV})$~\cite{Chung:2001cb}. However, the upper bounds $H_I \lesssim \mathcal{O}(10^{10})$~GeV (cf. Eq.~\eqref{hinf}) and $\Trh \lesssim 10^{11}$~GeV imply that it is not 
    robust for the present scenario.

	%%%%%%%%%%%%%%%%%%%%%%%%%%%%%%%%%%%
	\subsubsection{Inflaton Mediation}
    Alternatively, DM could also be produced by 2-to-2 scattering of SM particles, mediated by the $s$-channel exchange of inflatons. 
    The corresponding interaction rate density for $T \ll m_\phi$ is
    \begin{equation} 
		\gamma(T) \simeq \frac{y^2_{\chi}\, \lambda_{12}^2 }{2 \pi^5} \frac{T^6}{m_\phi^4}\,.
	\end{equation}
	%with $C \sim 4$.
	In appendix~\ref{app1} details about the derivation of the rate density are presented. 
    As in the present scenario $\Trh \ll m_\phi$ (cf. Fig.~\ref{fig:RH}), the DM abundance produced after reheating ($\Trh > T$) can be estimated with Eq.~\eqref{eq:BEafter}, and is given by
	\begin{equation} %\label{final_Yield}
	Y_0 \simeq \frac{135\,  y^2_{\chi}\, \lambda_{12}^2}{4 \pi^8\, \gss} \sqrt{\frac{10}{g_\star}}\, \frac{M_P\, \Trh}{m_\phi^4}\,.
	\end{equation}
	It is interesting to note that away from the instantaneous decay approximation for the inflaton, DM could also be produced by the freeze-in mechanism, between $\Tmax > T > \Trh$.
	However, it has been noticed that the boost for an inflaton scaling like nonrelativistic matter during reheating, it is only significant if the temperature dependence of the production rate density is high enough, $\gamma(T) \propto T^n$ with $n > 12$. In that case, the DM abundance is enhanced by a boost factor proportional to $(\Tmax/\Trh)^{n-12}$~\cite{Garcia:2017tuj, Bernal:2019mhf}.
	However, in the present scenario with $n = 6$, the difference between the standard UV freeze-in calculation differ only by an $\mathcal{O}(1)$ factor from calculations taking into account noninstantaneous reheating.
	Additionally, let us note that this production channel mediated by the exchange of inflatons turns out to be subdominant with respect to the direct decay of the inflaton, smaller by a factor $\frac{\Trh}{M_P} \left(\frac{\Trh}{m_\phi}\right)^3 \ll 1$.

    Finally, let us emphasize that for the previous analysis to be valid, the DM has to be out of chemical equilibrium with the SM bath.
    One needs to guarantee, therefore, that the interaction rate density $\gamma(T) \ll H(T)\, n_\text{eq}(T)$, where $n_\text{eq}(T)$ is the equilibrium DM number density.
    In our setup, for $T \leq \Tmax$, this condition is always satisfied.

	%%%%%%%%%%%%%%%%%%%%%%%%%%%%%%%%%%%%%%%%%%%%%%%%%%%%%%%%
	\section{Summary and Conclusions} \label{summary}
	%%%%%%%%%%%%%%%%%%%%%%%%%%%%%%%%%%%%%%%%%%%%%%%%%%%%%%%%
	In this paper, we have presented a minimal UV complete framework to embed cosmic inflation and dark matter (DM) via extending the standard model with two degrees of freedom: a real scalar inflaton and a fermionic DM. 
	For the inflaton potential, we take the most general polynomial of degree four, which admits a near inflection point to match CMB observations.
 
    This scenario produces a running of spectral index $\alpha \sim \mathcal{O}(10^{-3})$ and a tensor-to-scalar ratio  $r \simeq 7.09 \times 10^{-9} (\phi_0/M_P)^6$, with $3 \times 10^{-5}~M_P \lesssim \phi_0 \lesssim M_P$.
    Whereas the prediction for $r$ is beyond the future detection perspectives, $\alpha$ could be testable with next generation CMB experiments combined with small structure data.

    In the early universe, DM particles can be produced by a number of processes that counts the direct decay of the inflaton, the 2-to-2 scattering of standard model particles via the $s$-channel exchange of inflatons or gravitons, or purely gravitational interactions.
    However, due to the upper bounds on the reheating temperature (to guarantee the flatness of the inflaton potential), the pure gravitational channels are  {\it not} robust to produce enough DM, and  freeze-in  is  {\it always} subdominant compared to the direct decay. The viable parameter space to reproduce correct DM relics is shown in Fig.~\ref{fig:decay} by inflaton decay with branching ratio: $10^{-21} \lesssim \text{Br} \lesssim 10^{-4}$, where DM mass spans a large range, from the keV scale up to $\mathcal{O}(10^{11})$~GeV.
	
    Finally, we note that this model can be further extended to dynamically generate the baryon asymmetry of the universe. One can  consider the inflaton to be a complex field, %in which case the cubic term in the potential breaks CP symmetry, 
    so that such a model can naturally include the Affleck-Dine  mechanism for baryogenesis. However, this is beyond the scope of the present study and we leave it for future work.
	
	%%%%%%%%%%%%%%%%%%%%%%%%%%%%%%%%%%%%%%%%%%%%%%%%%%%%%%%%
	\section*{Acknowledgments}
	%%%%%%%%%%%%%%%%%%%%%%%%%%%%%%%%%%%%%%%%%%%%%%%%%%%%%%%%
    We are grateful to Manuel Drees for useful discussions and comments on the draft. NB received funding from Universidad Antonio Nariño grants 2019101 and 2019248, the Spanish FEDER/MCIU-AEI under grant FPA2017-84543-P, and the Patrimonio Autónomo - Fondo Nacional de Financiamiento para la Ciencia, la Tecnología y la Innovación Francisco José de Caldas (MinCiencias - Colombia) grant 80740-465-2020.
	This project has received funding /support from the European Union's Horizon 2020 research and innovation programme under the Marie Skłodowska-Curie grant agreement No 860881-HIDDeN. 
	
	\appendix
	\section{Interaction Rate} \label{app1}
    DM can be produced in the early universe via scattering of Higgs bosons mediated by the $s$-channel exchange of inflatons.
    The annihilation cross section for: $\bar{\chi} \chi \to H^{\dagger}H $  is given by
	\begin{equation} \label{crossection}
	\sigma(\mathfrak{s}) \simeq  \frac{1}{8\pi} \frac{y^2_{\chi}\, \lambda_{12}^2}{(\mathfrak{s} -m_\phi^2)^2 + m_\phi^2\, \Gamma^2}\,, 
	\end{equation}
	where $\sqrt{\mathfrak{s}}$ corresponds to the center-of-mass energy. We note that in the parameter space fulfilling all constraints mentioned in the text one has $\Gamma \ll m_\phi$, and therefore the narrow width approximation can be used. The corresponding interaction rate density  defined by $ \gamma \equiv  \sv n_{\text{eq}}^2$ and can be written as \cite{Bernal:2020fvw}
	\begin{equation}
	    \gamma(T) \simeq \frac{T}{8\pi^4} \int_{4 m_\chi^2}^\infty d\mathfrak{s}\, \mathfrak{s}^{3/2}\, \sigma(\mathfrak{s})\, K_1\left(\frac{\sqrt{\mathfrak{s}}}{T}\right)\,,
	\end{equation}
	where $K_1$ is the modified Bessel function of order 1.
    Several useful approximations can be implemented for different ranges of $T$, such that
	\begin{equation} \label{gamma}
		\gamma(T) \simeq \frac{y^2_{\chi}\, \lambda_{12}^2}{64 \pi^4} \times 
		\begin{cases}
			\frac{32}{\pi} \frac{T^6}{m_\phi^4} &\quad \text{ for } T \ll m_\phi\,,\\
			\frac{m_\phi^2\, T}{\Gamma}\, K_1\left(\frac{m_\phi}{T}\right) &\quad \text{ for } T \sim m_\phi\,,\\
			\frac{m_\phi\, T^2}{\Gamma} &\quad \text{ for } T \gg m_\phi\,.
		\end{cases}
	\end{equation}
	%

%%%%%%%%%%%%%%%%%%%%%%%%%%%%%%%%%%%%%%%%%%%%%%%
    \section{\boldmath Lyman-$\alpha$ Bound} \label{app2}
    Due to the large initial momentum of DM particles, they could have a  rather large free-streaming length; this  leads to a suppression on the structure formation at small scales.
    In our scenario here, DM has very suppressed interactions with the SM or with itself, so momentum of DM simply redshifts. And its present value $p_0$  is
    \begin{equation}
        p_0 = \frac{a_\text{in}}{a_0}\, p_\text{in} = \frac{a_\text{in}}{a_\text{eq}}\, \frac{\Omega_R}{\Omega_m}\, p_\text{in} = \left[\frac{\gss(T_\text{eq})}{\gss(\Trh)}\right]^{1/3}\frac{T_\text{eq}}{\Trh}\, \frac{\Omega_R}{\Omega_m}\, \frac{m_\phi}{2} \simeq 3 \times 10^{-14}\, \frac{m_\phi}{\Trh}~\text{GeV}\,,
    \end{equation}
    where $p_\text{in} \simeq m_\phi/2$ is the mean initial momentum at production (i.e., at $T = \Trh$), $T_\text{eq}$ and $a_\text{eq}$ correspond to the temperature and the scale factor at the matter-radiation equality, respectively.
    Additionally, we have used $T_\text{eq} \simeq 0.8$~eV, $\Omega_R \simeq 5.4 \times 10^{-5}$ and $\Omega_m \simeq 0.315$~\cite{Aghanim:2018eyx, Zyla:2020zbs}.
    A lower bound on the DM mass can be obtained from the upper bound on a typical velocity of warm dark matter (WDM) at present.
    Taking $v_\chi \lesssim 1.8 \times 10^{-8}$~\cite{Masina:2020xhk} for $m_{\text{WDM}} \gtrsim 3.5$~keV~\cite{Irsic:2017ixq}, one gets
    \begin{equation} \label{ly_mass}
        \frac{m_\chi}{\text{keV}} \gtrsim 2\, \frac{m_\phi}{\Trh}\,,
    \end{equation}
    which by using Eq.~\eqref{y0total}, implies an upper bound on the branching ratio
    \begin{equation} \label{br_bound}
        \text{Br}~\lesssim 10^{-4}.
    \end{equation}
    
	\bibliographystyle{JHEP}
	\bibliography{biblio}
\end{document}